# TARVOS – an Event-Based Simulator for Performance Analysis, Supporting MPLS, RSVP-TE, and Fast Recovery

Marcos Portnoi, *Member, IEEE*
Universidade Salvador
R. Ponciano de Oliveira 126, Rio Vermelho, Salvador, BA, Brazil. CEP 41950-275
mportnoi@ieee.org

Joberto S. B. Martins
Universidade Salvador, Computer Networks Research Group, Distance-Learning Research Group
R. dos Colibris 18, Salvador, BA. CEP 41370-410
joberto@unifacs.br

## ABSTRACT
This paper presents a new discrete event-based network simulator named TARVOS – Computer Networks Simulator, being designed as part of the first Author's Masters research and will provide support to simulating MPLS architecture, several RSVP-TE protocol functionalities and fast recovery in case of link failure. The tool is used in a case study, where the impact of a link failure on a VoIP application, within an MPLS domain network, is analyzed. The paper displays a preliminary research of six already available simulators and reasons why they were not adopted as tools for the Masters research. Then, it follows to describe the basics of TARVOS implementation and exhibits the case study simulated by this new tool.

## Categories and Subject Descriptors
I.6.2 [**Simulation and Modeling**]: Simulation Languages---event-based simulator.

## 1. INTRODUCTION
Services based on Web technology (Web services) are experiencing a significant growth. In the internet, for instance, Web 2.0 websites provide applications for the end-user such as spreadsheets, text editors, and schedule and appointment managers. In addition, telephony (Voice over IP or VoIP) and video tend to be massively used, thanks to the introduction of low-cost services such as Skype [24] and websites based on user-generated video content, of which YouTube [26] is the main example.

Many of these services require constraints regarding Quality of Service (QoS): VoIP applications show little tolerance to packet loss and delay; video streaming can sustain a certain loss, but it is very sensitive to delay and jitter. One of the causes of packet loss is network failure. A broken link will result in packets being discarded until a new route is made operational by the network. The overall process of failure recovering will cause extra delay. The Internet Protocol (IP) used today in the internet is robust and capable of restoring connectivity after several types of network elements malfunctions; it is, however, a best effort based protocol; therefore, it does not guarantee or provide any form of QoS. The time it takes IP to re-establish connectivity might not be within the boundaries required by end-user applications [21]. Moreover, conventional routing protocols used with IP (BGP, OSPF) in the internet, following the best effort philosophy, do not take into account, for their routing calculations and decisions, link capacity and traffic characteristics, resulting possibly in underutilized or over utilized paths, leading to congestion and packet loss and delay. Mechanisms need to be added to conventional TCP/IP to achieve control of traffic flows through a network in order to optimize performance and resource utilization (Traffic Engineering).

The MultiProtocol Label Switching (MPLS) [23] is an architecture that, along with a signaling protocol such as RSVP-TE [3], can be used as a tool to implement Traffic Engineering (TE) in networks. MPLS and RSVP-TE capabilities enable constraint routing, tunneling and mechanisms for fast rerouting and recovery. These may be formatted to guarantee levels of QoS for applications.

The first Author's Masters research intends to build a prototype for a computer networks simulation tool, and, in order to validate it, investigate the impact of MPLS and RSVP-TE on network performance, especially in case of link failure, from the point of view of a VoIP application. The methodology adopted was the use of the conceived network simulator to obtain the performance measures from a test topology, all presented in this paper.

The paper is formatted as follows: in order to justify construction of a new simulator, six available computer network simulators were analyzed in Section 2. Section 3 describes the innings and functionalities of TARVOS simulator. The case study or investigation is presented in Section 4, and Section 5 concludes the paper and suggests future work.

## 2. AVAILABLE NETWORK SIMULATORS
### 2.1 OPNET
The first simulator tested was OPNET [19], from Opnet Technologies, Inc. It provides several modules for network simulation comprising a vast universe of the protocols and network elements needed. The module for MPLS and RSVP-TE is available as a separate purchase from the standard commercial version.

OPNET is a commercial product and a license for its use is not available in the Authors' University. There is a free academic version, but it is limited and its documentation is yet poor, reasons that motivated building the new tool described here.

### 2.2 NS-2
NS-2 [18] is a discrete event simulator targeted at network research. It is open source, developed mainly by VINT project, Xerox PARC, UCB, USC/ISI, and contributions by several other researchers and users.

NS-2 is coded in C++ in a modular fashion. The user interfaces with the simulator using the object-oriented script language OTcl. It was conceived natively to run under Unix systems (including Linux), although it is possible to install it under Microsoft Windows [22].

MPLS and RSVP-TE are not available as standard libraries in NS-2. They were implemented through contributions from other researchers. The MNS (MPLS for Network Simulator) module was developed by Gaeil Ahn [9][10], its original location no longer being available in the internet. This module contains MPLS and CR-LDP, but not RSVP-TE. The MNS module was further extended by [4] and [1][7] to include RSVP-TE functionalities. These modules cannot be obtained directly from their authors' websites, but only through request by email or from users who already own the modules.

NS-2 learning curve is significantly steep. One has to know the script OTcl language and learn how to build scripts that interface with the simulation objects coded in C++. The available documentation is not written in a didactic style, making it difficult for the beginner to build initial simulations without investing a considerable amount of time in trial and error. The documentation is especially poor for the MPLS and RSVP-TE modules, requiring the user to read the source code in order to learn how to interface with it and detect the offered capabilities. It is open, but implementation of new functions or modifications demand studying large portions of the source code. Generation of results and statistics is not automatic. One has to build a trace file from the simulation and perform a post processing on the file, calculating the desired statistics, by means of a processing language such as awk. Simulations can easily produce very large trace files, demanding significant post processing times.

Due to those characteristics and to the fact that the main module needed for the simulations, MNS, is yet not fully supported, NS-2 also stimulated the devise of the new simulator.

## 2.3 CSIM19
CSIM19 [16] is a process-oriented, discrete event simulator available in either C, C++ or Java. It provides libraries that a program written in the same language can use in order to model a system and simulate it. It is a general simulator, not specific to computer networks, and it is commercial, bearing no free version. This alone discouraged its use in the work of the dissertation, and inspired the construction of an open tool that bore its good features.

## 2.4 Cnet v.2.0.10
Mainly developed for use in undergraduate computer networking courses, Cnet [15] is an event-driven simulator written in C and uses Tcl/Tk to implement its graphical interface, where the simulator shows a representation of the topology (topologies are constructed by means of *topology files*), and allows some attributes to be configured. The purpose of this simulator is to enable experimentation with networking protocols. IEEE 802.3 Ethernet and point-to-point WAN are built-in, and there is a mechanism to cause corruption or loss of data frames according to probabilities.

Cnet manual states that the simulator runs only on Linux/Unix platforms. It is not compatible with Microsoft Windows systems or Apple Macintosh, and does not implement MPLS or RSVP-TE protocols. Although the source code for the simulator is freely available, it is not thoroughly documented, and it natively generates only basic statistics. The focus of this simulator is protocol building and implementation, not QoS analysis, and, together with platform limitations and lack of MPLS and RSVP-TE protocols, it was another strong motivator for the adoption of a self-made simulator for the research studies.

## 2.5 J-SIM
There are a number of simulators named J-Sim available through a simple internet web search. This J-Sim [12] is an open source, component-based simulator written in Java. It provides MPLS support through a third-party extension [13], but it does not include the RSVP-TE signaling protocol. The documentation is available from the simulator's website, and it does include good descriptions of native code implementation, the philosophy behind the simulator and some tutorials and guides for new implementations.

Installation of the simulator, as it seems usual with Java applications, requires setting environment variables and compiling the source codes with third-party tools more common in Linux/Unix platforms, and then applying patches needed by the extensions such as MPLS. J-Sim is a dual language environment, where the user manipulates classes written in Java using Tcl scripts, much resembling NS-2. This poses the same problems related to NS-2, i.e., the need to know both Java and Tcl in order to use the simulator and implement non-existent characteristics, and, for the same reasons, it has inspired the development of TARVOS.

## 2.6 OMNET++
This simulator is a discrete event environment programmed in C++, making up modules or components that are then assembled into models, using an internal language called NED (*NEtwork Description*) [25]. It is designed primarily to use in simulation of computer networks, but it admits being able to support simulation of queuing networks and other systems.

A network simulation is achieved using a model, called INET Framework, available along with several other models at the simulator website. This model was first implemented by Xuan Thang Nguyen, but the original website of this implementation is no longer reachable. Its documentation indicates support of MPLS forwarding, LDP and RSVP-TE. It is not clear, and the extensive documentation does not mention it, whether the RSVP-TE component includes Rapid Recovery or Failure Recovery.

Usage of Omnet is not straightforward. It comprises several windows and menus that require walking through a large web-based documentation, tutorials and demos in order to begin building simulation topologies. Its characteristics of being programmed in C++ and using a second, internal language (NED) to build topologies, its vast code and not clearly supporting Failure Recovery, fostered the creation of TARVOS.

## 3. BUILDING TARVOS
For the intents of the research for the Masters dissertation, a flexible, configurable, straightforward, open and customizable simulator was needed. Moreover, one that could produce results with a minimum of post processing of trace files, preferably results that could be programmed and calculated during the simulation. Finally, one that provided MPLS and RSVP-TE functionalities for testing failure recovery. A new simulator was then prototyped: it was named TARVOS *Computer Networks Simulator*, or TARVOS for short.

## 3.1 Characteristics
TARVOS is a discrete event-based simulator, coded entirely in C. It was developed as an extension for C, containing functions and structures that model queuing systems and, on top of those, network elements. It consists of three main elements: the *kernel*, the *shell 1* and the *shell 2*, as portrayed in Figure 1.

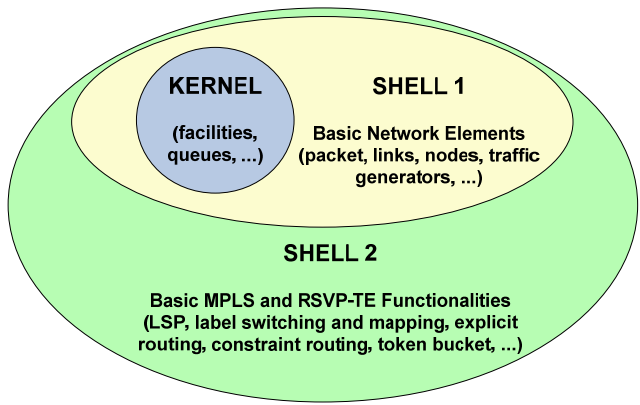

Figure 1: TARVOS Construction.

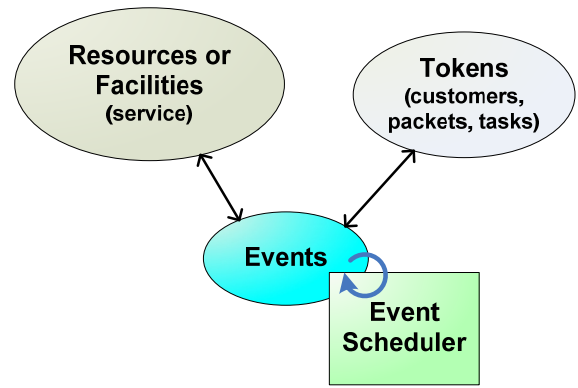

Figure 2: Relationship of system entities, as seen by TARVOS kernel.

### 3.1.1 The Kernel

The *kernel* is a discrete event-based queuing systems simulator based on the SMPL simulator [14]. The designer of this kernel [6] conceived it by rewriting SMPL using dynamic allocation, structures and pointers, allowing a complex structure to be passed through events, whereas the original SMPL only allowed a single integer to be passed. The kernel implements basic queuing system elements, such as resources and facilities that provide the service; priority and FIFO queues for the facilities; functions and data structures for event manipulation; and statistic functions, random number generators and random variate generators. The kernel was extended and modified to support down (no operational) servers, a well-behaved priority queuing, more statistics, and random variate generators such as Pareto and Exponential On/Off.

The kernel sees a system as a combination of three main components or entities: *resources*, *tokens* and *events* (Figure 2).

A system is composed by an interconnected collection of *resources* [14]. Theses resources, in real life, may be routers, processors, bank tellers, web servers. Basically, anything that deliver some sort of *service*. The resources are called *facilities* in TARVOS, and have a series of functions related to them: definition, reservation, release, preemption and status.

A facility is defined by configuring its name and its number of servers. When a facility is required to perform a service on a customer, a *reservation* is requested for it. When the service is finished, the facility is *released*. The busy condition of a facility is reported by its *status*.

The *tokens* represent the active entities of the system; they can model customers, tasks, network packets, people, automobiles, bytes, etc. The dynamic behavior of a system is modeled by the flow of tokens through the collection of resources or facilities. In TARVOS, a token is a data structure. If this data structure models a network packet, then it contains typical packet information, such as source node, destination node, packet ID, MPLS label, priority, etc.

The *facilities*, therefore, perform service on the *tokens*. A typical router, modeled as a facility with one server, would receive, transmit or enqueue a packet (modeled as a token). A network link is modeled as a facility capable of transmitting the packet at a given speed (calculated from the link's bandwidth), and then propagating the packet through the medium. The last is modeled as a constant delay, its value dependent on the medium type.

In the process of a simulation, if a token needs to be serviced by a facility, then a *request* for service is made: if the facility has free servers, the token is put into service, and the *time* the service must be concluded is *scheduled*. This is called an *event*. At that time, the server, within the facility, that held the token is *released*. The token then must continue its way through the system, requesting service at other facilities.

If, when requesting service, a token finds a facility with all busy servers, then this token is *enqueued* at the facility's queue. TARVOS provides a Priority Queue (PQ) for each facility. The priority is an integer. Higher numbers mean higher priority. This priority is specified when requesting service at a facility, and can be contained within the token's data structure. When the facility is released, it collects the first token in its queue and puts it in service. (Here, it is interesting to mention, when enqueing a token, the simulator also records the service time for that token, so as when the token is dequeued and put into service, the simulator knows when to schedule the release.)

A facility can also be *preempted*, i.e., when a token with a certain priority requests service and the facility is busy, then a token in service, with a lower priority than the requesting token, will have its service interrupted, its remaining service time computed, and will be enqueued before tokens of the same priority. The requesting token with higher priority will then take its place in the now vacant server. When this token finishes service, the server is released and the simulator dequeues the token at the head of the queue. If this is the preempted token, then the server reserved for it has its release time scheduled for the remaining service time of the preempted token. If, when requesting a preemption in a busy facility, no tokens with lower priority than the preempting token are found, then the preempting token is enqueued in the same way as a normal, non-preemptive service request.

An *event* is any change of state in the system. For instance, the arrival of a new token, a service request for a token, a facility release, a dequeue. Events are identified in the simulator by a number, the time instant when this event must occur and the token related to it. The *event scheduler* manages events by organizing them into an *event chain*, a double-linked list ordered by time. Events are continuously generated in the simulation, until a certain stop criteria is met. Three main functions are related to events: *schedule* an event (or put the event into the event chain); *cause* an event (or retrieve the next event from the event chain), and *cancel* an event (remove it from the event chain and discard it).

Tokens move between resources through the scheduling of events in time. Interconnection between resources or facilities is not explicit for the kernel. This interconnection is implied from the routing of tokens through the facilities; the routing, in turn, is defined by the processing of each type of event. This processing, from the kernel's point of view, depends entirely on the user, who will code it in C for each event type.

### 3.1.2 Shell 1

The *shell 1* provides libraries that implement basic network elements, such as packets, simplex and duplex links, nodes, static routing, link failures, and traffic generators (exponential On/Off and CBR). It also comprises data structures that facilitate calculation of performance measurements and statistics, and functions to generate traces. Shell 1 functions use mainly kernel functions to perform their duties.

The *nodes* are modeled as entities capable of receiving packets from links, forwarding packets to links, making decisions about paths and routes, discarding packets, and collect statistics such as delay and jitter. In short, they behave much as network routers.

*Links* entities connect two nodes in one way only (simplex links) or two ways (duplex links, which are in fact two simplex links). They are modeled, in TARVOS's kernel point of view, as facilities with one server per simplex link. The service provided is transporting packets from one end of the link to the other end, i.e., from one node to another. This is done in two steps: upon being requested, the links first transmits the packet at full bandwidth speed (before actually being transmitted, packets might be policed by the simulator's Policer, based on a token bucket algorithm); the time it takes to transmit a packet is a relationship between packet size and bandwidth speed. If transmission is successful, the packet is now considered to be into the link medium (the wire, for instance). Then, second, the packet propagates through the medium until the destination node. This propagation is a simple fixed delay, defined by the user. Thus, in fact, a network link is modeled in TARVOS as a facility with one server and a priority queue, chained to an infinite-capacity server, which represents the link medium.

*Packets*, as mentioned before, are data structures in the simulator. Packet size in bytes, source and destination node, ID number, MPLS label, message ID and type, and explicit route object are some information stored in those data structures.

*Traffic generators* are entities connected to nodes that generate packets according to probability distributions. These distributions, in turn, model real network traffic. TARVOS includes a CBR (*Constant Bit Rate*) generator, Exponential (or Poisson) generator, Exponential On/Off and Pareto. The Exponential On/Off can be used to model VoIP applications.

### 3.1.3 Shell 2

The *shell 2* consists of libraries supplying basic MPLS and RSVP-TE [3] control plane, forwarding and signaling functionalities, including label switching, primary- and backup-LSP creation, explicit and constraint routing, traffic policer, soft-state maintenance, RSVP-TE PATH, RESV and HELLO messages, and mechanisms for failure detection and recovery.

In order to establish an LSP tunnel between two nodes, called the ingress LER (*Label Edge Router*) and egress LER, the user invokes a function, specifying the constraints the tunnel must meet and the explicit route. The simulator creates a PATH_LABEL_REQUEST message, encapsulates it into a packet, and sends it to the egress router. If each hop is capable if meeting the constraints, the PATH message reaches the destination egress LER; this LER creates a RESV_LABEL_MAPPING message, encapsulated into a packet, and sends it back to the ingress LER. This message confirms reservations made along the path and performs the label mappings, by means of populating a Label Information Base (LIB) table [2]. Once the LSP tunnel is fully set up, a traffic policer may be activated in order to force packets to comply with the LSP constraints.

The user creates backup LSPs, in a one-to-one method [20], invoking a specific function, giving the primary LSP number, beginning and ending nodes (named Merge Points), and the explicit route as parameters. The simulator composes a PATH_DETOUR message, which traverses the explicit route, pre-reserving resources if available. If resources are already reserved for the primary LSP in any part of the path, the simulator does not perform a new reservation; this way, the reserved resources are shared between the primary and backup LSPs. Once the PATH_DETOUR reaches de destination node (which signals that the backup path meets the same constraints as the primary LSP), a RESV message is sent back to the beginning Merge Point, confirming the resource reservations.

The simulator maintains RSVP soft state by means of PATH_REFRESH and RESV_REFRESH messages, generated in a timely basis (the generation period is user configurable). HELLO messages are also supported [3].

Failure detection and fast recovery are triggered mainly by timeouts in the LSPs soft state refreshes and errors in HELLO messages. When a node is unreachable, mainly due to a failed link, LSPs that traverse that link will time out. Nodes that send HELLO messages to other nodes through that link will not receive HELLO ACKs. Fast recovery [20] will attempt to find a detour LSP for timed out LSPs around the failed links. This is done at node level, so no signaling is actually needed to inform other nodes or applications of a recovery being made.

## 3.2 Preparing a Simulation

The user constructs a simulation by writing a C program, composed of at least the *main* function. This program will use the functions and structures provided by TARVOS to model the network topology and handle a series of events. The following paragraphs describe the steps that need to be taken in the user program.

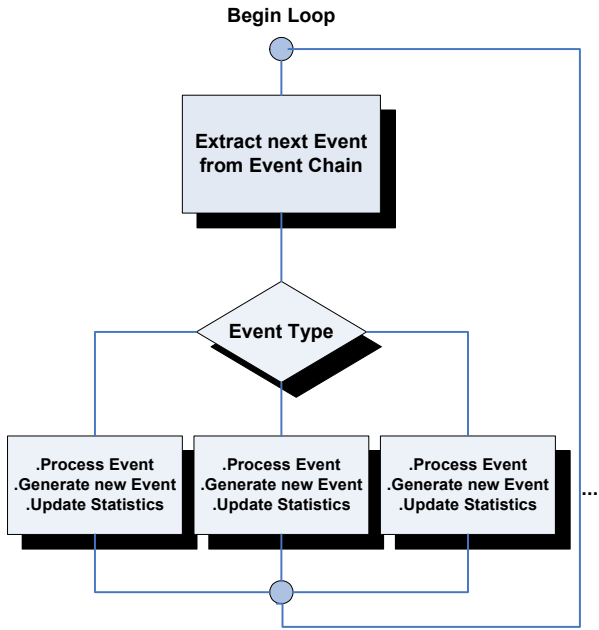

**Figure 3: Simulation main loop.**

First, the user builds the topology model by creating nodes and links that connect the nodes, and attaching traffic generators to desired nodes. The shell functions *createTrafficSource*, *createDuplexLink*, and *createNode* provide the means to create those elements. Explicit routes and other structures to collect statistics can also be defined here.

Second, the user sets up primary LSPs (by calling the shell function *setLSP*), backup LSPs (calling shell function *setBackupLSP*), and schedule initial events and timers, such as the end of the simulation and the start of the traffic generators.

Third, the program enters a loop, retrieving a new event from the event chain and treating this event accordingly, until the scheduled end of the simulation is reached. Typically, this part of the program is a *switch-case* structure within a *while* loop (Figure 3). Each *case* treats a specific event by calling shell functions and, if needed, generating other events. The next section brings an overview of the typical events in a TARVOS simulation.

Fourth, after the end of the loop, the program collects data from data structures, and performs and records the desired statistics calculations (for instance, delay, jitter, packet loss).

### 3.3 Events

Literature usually states that one problem with event-based simulators is their limited scalability; they are said to be suitable for small and middle scale simulations. In TARVOS, effort was done so as to limit the number of events the user program must predict and handle. Several functions were created to mimic network operations and to keep the number of steps a user must process at each event at a minimum. A typical user simulation will handle mainly ten different types of events: *arrival of packet from traffic source*, *link transmit request, propagate packet through link, arrival of packet at node, arrival of control message, refresh LSP states, generate HELLO message, timeout trigger, start traffic generator*, and *end simulation*. TARVOS functions are responsible to break the events to the level of specific nodes, links, or traffic generators.

In the *arrival of packet from traffic source* event, the user program must schedule a *link transmit request* event for the packet (by calling a TARVOS kernel function), schedule a new arrival from the same traffic source (by calling a TARVOS shell function), and trigger the reception of the packet by the current node (by calling a TARVOS shell function). Here, the traffic policer can also be invoked in order to guarantee conformity to constraints. This event represents the entry of a packet into an MPLS domain.

In *link transmit request*, the user program calls a shell function to decide the next node the packet should be sent, either by explicit routing, label switching or static routing. Then, calls another shell function that transmits the packet (or puts it into the queue, if the resource is busy) and schedules a *propagate packet through link* event.

The *propagate packet through link* event, the end of the packet transmission should be scheduled, and the propagation itself is activated. Then, an *arrival of packet at node* event is scheduled. All of this is achieved by calling two shell functions.

*Arrival of packet at node* invokes the reception of the packet by the current node. If the current node is not the destination of the packet, and the packet was not discarded, then a *link transmit request* event is scheduled.

The event *arrival of control message* should do the same as *arrival of packet at node*, i.e., schedule a *link transmit request*. It is the initial event called when a control message (such as PATH or RESV) is first created; the entrance of a control message into the MPLS domain.

In *refresh LSP states*, the user invokes the shell function that triggers the refresh of all LSPs. It is interesting to notice that synchronization in the creation of PATH_REFRESH messages is avoided in the simulator, as instructed in [5]. This event must also be re-scheduled.

Now, *generate HELLO message* triggers generation of HELLO messages by all nodes, using algorithms to avoid synchronization. Next, the same event is re-scheduled.

*Timeout trigger*, as the name suggests, triggers timeout verification a series of timers, including LSP states and reservations, and reception of control messages ACKs. The same event is also re-scheduled.

*Start traffic generator* is an event that can be used to start traffic generators at a specific time, by invoking TARVOS shell traffic generators functions.

Finally, *end simulation* breaks the *while* loop, allowing the simulation to end.

## 4. CASE STUDY: Impact of Fast Recovery on a VoIP Application

To investigate and validate the simulator, a test topology was designed, as seen in Figure 4.

The topology is composed of 10 nodes. One primary, or protected, LSP runs through nodes 1-2-3-4-5-6. One detour or backup LSP runs through nodes 2-7-8-9-4-5-6 (shown in Figure 4 as "Detour LSP"). Two other backup LSPs have paths through nodes 3-8-9-5-6 and 3-10-5-6. On node 1, an Exponential On/Off traffic generator is attached, modeling a VoIP application, sending traffic to node 6.

It generates 512-byte packets at 64Kbits per second (the speed of a PCM codec) during the ON periods.

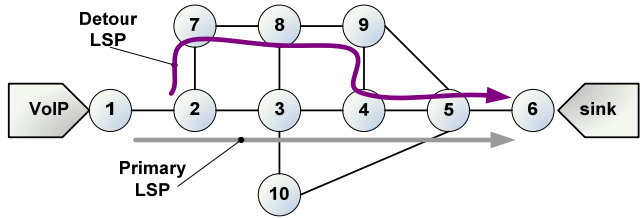

**Figure 4: Test topology for case study.**

The ON periods are distributed exponentially with mean 1.2 seconds, and the OFF periods are distributed exponentially with mean 0.8 seconds (these means are mentioned in literature as empirically satisfactory to model VoIP with an exponential On/Off generator). All links are of 10Mbit bandwidth, 10ms propagation delay. No traffic policer is in effect in this case study. The exponential On/Off generator is started 5 seconds after the simulation is initialized, to allow for the setting of all LSPs.

PATH messages have a fixed length of 120 bytes [8] and RSVP states are refreshed every 30 seconds (i.e., 120-byte PATH_REFRESH messages are generated for each LSP every 30 seconds). The timeout for RSVP states is 90 seconds. HELLO messages are generated in every node every 5 ms, and these messages are 20-byte long; the timeout for reception of HELLO ACKs is 17.5 ms [3]. The simulator performs a timeout check for states and control messages every 5 ms. The stopping criteria for the simulation is 50 seconds (simulated time).

At 10.029 seconds from the beginning of the simulation, the link connecting nodes 2 and 3 is scheduled to fail (this value was chosen to match an ON time of the VoIP generator). This link is brought up again at 15 seconds from start. When the link fails, traffic from VoIP running through nodes 1-2-3-4-5-6 is disrupted. The fast recovery mechanism of the MPLS domain is activated when HELLO messages from node 2 to node 3 are lost and time out. Locally, node 2 begins searching for primary LSPs established on the failed link. For each one found, node 2 searches for correspondent backup LSPs. When a backup LSP is found, node 2 updates its LIB entry, so packets label switched that would go through the failed link, now will traverse the link connecting nodes 2 and 7. The remaining of the label switched path will conduct packets from the current LSP through nodes 7-8-9-4-5 and 6.

When the functionalities of the failed link are resumed at 15 seconds simulation time, the HELLO messages acknowledge this. The LSP is not brought back to its original path, though, since this capability is not programmed into the simulator.

Two measurements were collected for this case study: Application Delay (difference between the time a packet was sent and when it was received) shown at node 6 (in seconds), and Application Jitter (variation of the delay between two packets) at node 6 (in seconds). In other words, the delay and jitter for packets exclusive from the VoIP generator were calculated and recorded at their destination at node 6.

In Figure 5, delay is recorded as being approximately 0.052 s or 52 ms before link failure. This figure is the approximate sum of all link propagation delays (a total of 5 links) plus transmission times. Notice that there are no competing traffic for the VoIP generator. During ON periods, a packet is received every 0.064 seconds, which is coherent with 512-byte packets being generated at 64Kbps (intergeneration time = (512 * 8) / 64000) = 0.064 s).

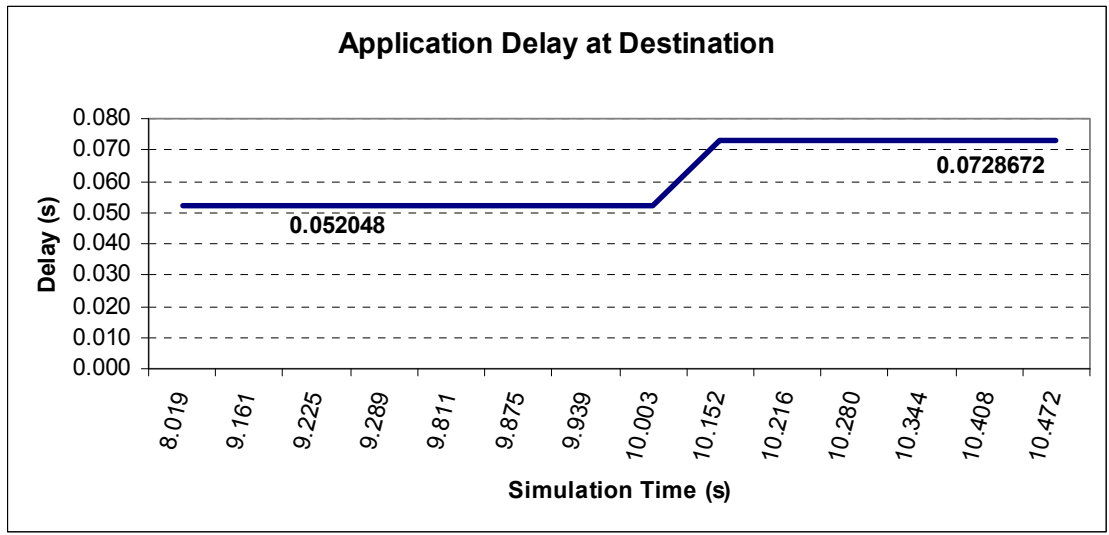

**Figure 5: Application Delay measured at destination (seconds).**

In this scenario depicted in Figure 5, node 6 will perceive problems with communication from the VoIP application after the packet received at 10.003 seconds. The next packet should arrive at 10.067; a new packet is only received, though, at 10.152 seconds, representing an extra delay of 85 ms for this packet. Henceforth, interarrival times resume the expected value of 0.064 seconds during ON periods. Between 10.003 and 10.152, a fast recovery was performed locally in node 2, switching traffic from the original path to the backup path. This complete backup path traversing nodes 1-2-7-8-9-4-5-6 is longer than the primary one, containing two more links of 10Mbps and 10 ms of propagation delay. This reflects on the new delay figure, now of 72.9 ms approximately (the new figure contains the extra link delays and transmission times). The simulator also reported that 5 packets were dropped between nodes 2 and 3 at the moment of link failure, comprising control messages and application packets.

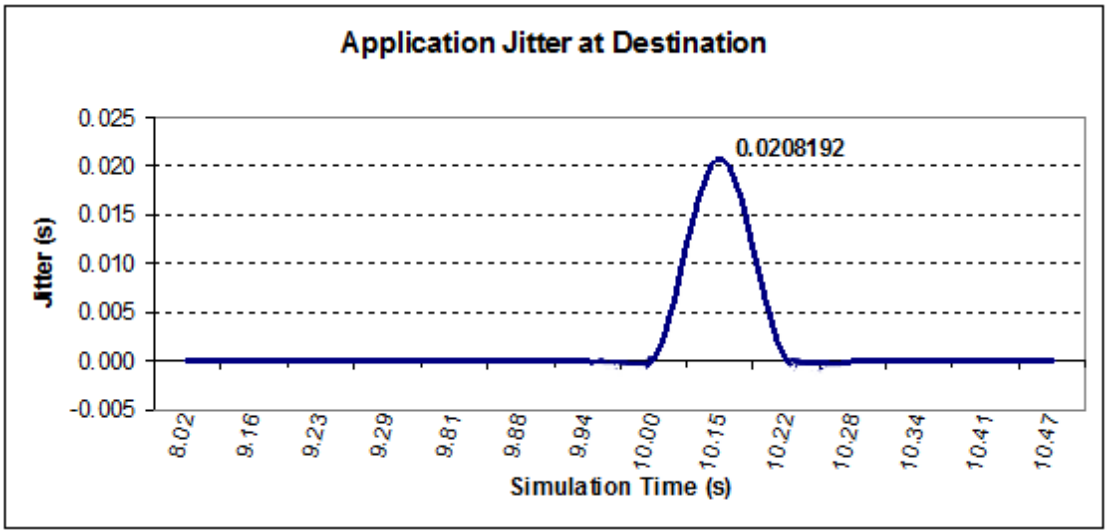

**Figure 6: Application Jitter measured at destination (seconds).**

In Figure 6, a measurement of the jitter in application packets shows basically no jitter moments before link failure. This is understandable, since the VoIP application is the only traffic in the network, not competing with any other traffic but the periodic generation of PATH, RESV and HELLO messages from the nodes. At the exact time the first packet is received at node 6, after link failure, the jitter is recorded at 20.8 ms approximately. Henceforth, the jitter is again brought to zero. The figure of 20.8 ms is not coincidentally the difference between the delay before failure (52 ms) and the delay after failure (72.8 ms).

### 4.1 Impact on Quality of Service

For VoIP applications, the G.144 recommendations from ITU-T [11] states that a delay from 0 to 150 ms is acceptable. For jitter values, one guideline [17] suggests that values inferior to 40 ms are not perceivable. Values from 40 to 75 ms are still of good quality, but noticeable. Values above 75 ms would be not acceptable.

In the case study, delays are always inferior to ITU-T recommendations. This is, of course, a simple testbed with no concurrent traffic. Jitter is measured at zero throughout most of the simulation, but when failure and fast recovery occur. Its value of 20.8 ms can be considered undetectable to a user, and it is important to notice it is present only once. The jitter rapidly resumes its average value of zero. Thus, it is a jitter caused by loss of packets due to link failure and extra delay caused by a longer path, traversed by the backup LSP. As soon as local fast recovery is completed (which happens almost instantly, since it is dependent only on the processing power of the local router), VoIP traffic resumes without further disruption.

### 5. CONCLUSION AND FUTURE WORK

This paper intends to present a new simulation tool for computer networks, named TARVOS. The contribution of this simulator to the research and academic community can be listed in the following points. The tool will be offered as open source. It can be fully customized and provides a high level of control, for the user, of the simulation, since the user programs the flow of events and is able to follow exactly the processes taken at each event. The use of the C language (and no other script language) makes it relatively easy to understand implementation of the simulator functions (by reading the source code) and for building simulation models. By means of a template, one can rapidly construct a *main* function to simulate a network topology and collect several statistics.

A case study is depicted, where the impact of MPLS fast recovery after a link failure is analyzed on a VoIP application. The simulated case study showed the behavior of MPLS fast recovery and how a VoIP traffic if influenced, from the traffic destination point of view. The simulation demonstrated that the link failure caused a disruption in traffic, visible specially in the jitter recordings. Fast recovery, however, limited this disruption to a very brief moment, allowing traffic to resume quickly.

Listed, as prospective work: the programming of TCP protocol simulation into TARVOS; inclusion of self-similar traffic generators, for modeling video applications; multiple queues for one facility and other queue disciplines, such as WFQ (Weighted Fair Queueing), WRR (Weighted Round Robin), RR (Round Robin); rebuild data structures to allow multiple simulations and automatic calculation of confidence intervals; include routing protocols (OSPF, for instance); and, finally, producing a complete documentation for the simulation tool.